\documentclass[fleqn,usenatbib]{mnras}

\usepackage{newtxtext,newtxmath}

\usepackage[T1]{fontenc}

\DeclareRobustCommand{\VAN}[3]{#2}
\let\VANthebibliography\thebibliography
\def\thebibliography{\DeclareRobustCommand{\VAN}[3]{##3}\VANthebibliography}

\usepackage{graphicx}	
\usepackage{amsmath}	

\newcommand{\coralie}{{CORALIE}}
\newcommand{\harps}{{HARPS}}

\newcommand{\teff}{{T$_{\rm eff}$}}
\newcommand{\logg}{{$\log$ g}}
\newcommand{\feh}{[Fe/H]}
\newcommand{\vsini}{$V\sin i$}

\pdfoutput=1

\title[EBLM X. Fully convective M-dwarfs with K2]{The EBLM project X. Benchmark masses, radii and temperatures for two fully convective M-dwarfs using K2}

\author[Alison Duck et al.]{
Alison Duck$^{1*}$$^{\href{https://orcid.org/0000-0002-4531-6899}{\includegraphics[scale=0.5]{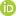}}}$,
David V. Martin$^{1,2}$ $^{\href{https://orcid.org/0000-0002-7595-6360}{\includegraphics[scale=0.5]{orcid.jpg}}}$,
Sam Gill$^{3}$$^{\href{https://orcid.org/0000-0002-4259-0155}{\includegraphics[scale=0.5]{orcid.jpg}}}$,
Tayt Armitage$^{1}$,
Romy Rodr\'iguez Mart\'inez$^{1}$$^{\href{https://orcid.org/0000-0003-1445-9923}{\includegraphics[scale=0.5]{orcid.jpg}}}$, \newauthor
Pierre F. L. Maxted$^{4}$ $^{\href{https://orcid.org/0000-0003-3794-1317}{\includegraphics[scale=0.5]{orcid.jpg}}}$,
Daniel Sebastian$^{5}$ $^{\href{https://orcid.org/0000-0002-2214-9258}{\includegraphics[scale=0.5]{orcid.jpg}}}$,
Ritika Sethi$^{1}$, 
Matthew I. Swayne$^{4}$ $^{\href{https://orcid.org/0000-0002-2609-3159}{\includegraphics[scale=0.5]{orcid.jpg}}}$, \newauthor
Andrew Collier Cameron$^{7}$ $^{\href{https://orcid.org/0000-0002-8863-7828}{\includegraphics[scale=0.5]{orcid.jpg}}}$,
Georgina Dransfield$^{5}$ $^{\href{https://orcid.org/0000-0002-3937-630X}{\includegraphics[scale=0.5]{orcid.jpg}}}$,
B. Scott Gaudi$^{1}$ $^{\href{https://orcid.org/0000-0003-0395-9869}{\includegraphics[scale=0.5]{orcid.jpg}}}$,
Michael Gillon$^{8}$ $^{\href{https://orcid.org/0000-0003-1462-7739}{\includegraphics[scale=0.5]{orcid.jpg}}}$,
Coel Hellier$^{4}$,\newauthor
Vedad Kunovac$^{9}$ $^{\href{https://orcid.org/0000-0001-9419-3736}{\includegraphics[scale=0.5]{orcid.jpg}}}$,
Christophe Lovis$^{10}$ $^{\href{https://orcid.org/0000-0001-7120-5837}{\includegraphics[scale=0.5]{orcid.jpg}}}$, 
James McCormac$^{3}$,
Francesco A. Pepe$^{10}$,
Don Pollacco$^{3}$,\newauthor
Lalitha Sairam$^{5}$ $^{\href{https://orcid.org/0000-0001-8102-3033}{\includegraphics[scale=0.5]{orcid.jpg}}}$,
Alexandre Santerne$^{11}$,
Damien Ségransan$^{10}$ $^{\href{https://orcid.org/0000-0003-2355-8034}{\includegraphics[scale=0.5]{orcid.jpg}}}$,
Matthew R. Standing$^{5}$ $^{\href{https://orcid.org/0000-0002-7608-8905}{\includegraphics[scale=0.5]{orcid.jpg}}}$, \newauthor
John Southworth$^{4}$, 
Amaury H. M. J. Triaud$^{5}$ $^{\href{https://orcid.org/0000-0002-5510-8751}{\includegraphics[scale=0.5]{orcid.jpg}}}$,
Stephane Udry$^{10}$ $^{\href{https://orcid.org/0000-0001-7576-6236}{\includegraphics[scale=0.5]{orcid.jpg}}}$\\
*duck.18@buckeyemail.osu.edu \\
Affiliations are listed at the end of the paper.
}

\date{MNRAS Submission}

\pubyear{2022}

\begin{document}
\label{firstpage}
\pagerange{\pageref{firstpage}--\pageref{lastpage}}
\maketitle

\begin{abstract}

M-dwarfs are the most abundant stars in the galaxy and popular targets for exoplanet searches. However, their intrinsic faintness and complex spectra inhibit precise characterisation. We only know of dozens of M-dwarfs with fundamental parameters of mass, radius and effective temperature characterised to better than a few per cent. Eclipsing binaries remain the most robust means of stellar characterisation. Here we present two targets from the Eclipsing Binary Low Mass (EBLM) survey that were observed with K2: EBLM J0055-00 and EBLM J2217-04. Combined with HARPS and CORALIE spectroscopy, we measure M-dwarf masses with precisions better than $5\%$, radii better than $3\%$ and effective temperatures on order $1\%$. However, our fits require invoking a model to derive parameters for the primary star and fitting the M-dwarf using the transit and radial velocity observations. By investigating three popular stellar models, we determine that the model uncertainty in the primary star is of similar magnitude to the statistical uncertainty in the model fits of the secondary M-dwarf. Therefore, whilst these can be considered benchmark M-dwarfs, we caution the community to consider model uncertainty when pushing the limits of precise stellar characterisation.
\end{abstract}




\section{Introduction}

M-dwarfs are the smallest and most common stars in the galaxy. Low-mass stars have become a popular target of exoplanet searches for several reasons. A low stellar temperature implies a habitable zone that is much closer to the star than for a solar analogue.  This amplifies the transit probability and radial velocity amplitude, and shortens the observing time-span needed to measure these signals. The occurrence rate of terrestrial planets is also known to be higher around M-dwarfs than K- and G-dwarfs \citep{Dressing2015,Hardegree2019}. Although the question of habitability around M-dwarfs is complex, including the potentially detrimental effects of flares and coronal mass ejections \citep{Sheilds2016, Gunther2020}, it is undoubtable that there has been significant investment in that field.

These attributes have led to a significant community push towards characterizing M-dwarfs. This includes transit surveys (e.g. MEARTH \citealt{Berta2013}, TRAPPIST \citealt{Gillon2017}, SPECULOOS \citealt{Sebastian2021}, and, to a certain extent, TESS \citealt{Ricker2014}) and the latest spectrographs with a redder wavelength coverage (e.g. NIRPS \citealt{Wildi2017}, CARMENES \citealt{Quirrenbach2014}, SPIRou \citealt{Thibault2012}, and HPF \citealt{Mahadevan2012}). A significant portion of JWST's exoplanet science will be dedicated to M-dwarfs, which are likely the only stars for which transmission spectroscopy of the atmospheres of small habitable-zone exoplanets will be possible \citep{Wunderlick2019,Phillips2021}. The upcoming ARIEL \citep{Pascale2018} and TWINKLE (\citealt{Edwards2019},\citealt{Phillips2022arXiv}) missions, dedicated to transmission spectroscopy,  will have a significant focus on M-dwarfs.

Observational constraints of exoplanets in these systems depend on our constraints on the host star. For M-dwarfs the stellar dimensions are less well-determined than for Sun-like stars. This is because the number of well-characterised M-dwarfs is small. Eclipsing binaries (EBs) are the classic means of calibrating precise stellar parameters, and we only know of dozens of EBs containing a fully-convective star ($M\lessapprox0.35M_\odot$) with mass and radius measured to a precision better than $5\%$. Furthermore, within this small sample of measurements there are disagreements with predictions from stellar models. Across all M-dwarf spectral types we have seen radii $\sim$5\% higher than expected \citep{Chabrier2000,Torres2014,Morrell2019}, commonly referred to as the ``radius inflation problem''. This problem is equally evident in single M-dwarfs and in EBs \citep{2013ApJ...776...87S}. There have also been discrepancies with respect to M-dwarf effective temperatures, with outliers appearing in various studies \citep{Ofir2012,MaqueoChew2014,Chaturvedi2018}, although they were refuted upon later examination (\citealt{Swayne2020}, Martin et al., under rev..). Overall, these M-dwarf observational discrepancies not only inhibit our ability to make precise exoplanet measurements, but they confound theories of stellar structure and evolution.

To help solve these problems we created the EBLM (Eclipsing Binary Low Mass) survey \citep{Triaud2013}. The survey is constructed of unequal mass, single-lined spectroscopic eclipsing binaries discovered with WASP photometry \citep{Pollacco2006}. They typically consist of F/G/K + M stars. One of the aims of the EBLM project is to empirically calibrate the mass-radius-temperature-metallicity relation for M-dwarfs. \citet{Triaud2017} presented the spectroscopic orbits for 118 EBLM systems. Over a dozen EBLM targets have received follow-up photometry (e.g. CHEOPS, TRAPPIST, EulerCam) to improve upon the coarse WASP photometry used for the initial detection \citep{vonBoetticher2017,Gill2019,vonBoetticher2019,Swayne2021}. TESS will provide photometry for the majority of the sample, and a subset of the available TESS data has already been analysed for a few targets \citep{hodzic2020,Swayne2021}. 

In this paper we analyse two EBLM targets, EBLM J0055-00 and EBLM J2217-04, which were observed by the K2 mission \citep{Howell2014}. This was the re-configured continuation of the Kepler mission \citep{Borucky2010} after failure of two of the four  reaction wheels. Despite the reduced pointing stability, the K2 photometry for these two EBLMs is of exceptional quality, superior to almost the entire EBLM sample and making these some of the best-characterised M-dwarfs known. Not only can we measure the radius to an exquisite precision, but there are clear secondary eclipses, and hence we can measure M-dwarf effective temperatures. In addition, EBLM systems are observed as part of the CHEOPS guaranteed-time observing program \citep{Swayne2021}.

Since the EBLM survey consists solely of small mass ratio binaries, the secondary star is typically more than 5 magnitudes fainter. This means that spectroscopically the secondary star is essentially invisible, making it a single-lined spectroscopic binary (SB1). The advantage of an SB1 is that we may achieve a radial velocity precision comparable to that around single stars of a similar brightness ($\sim1-2$ m/s), as demonstrated in the BEBOP (Binaries Escorted By Orbiting Planets)  survey for circumbinary planets \citep{Martin2019,Standing2022,Triaud2022}. 

The disadvantage of SB1's is that we do not measure $M_{\rm A}$ and $M_{\rm B}$ individually, but rather we measure the mass function $f_{\rm m} = M_{\rm B}^3/(M_{\rm A}+M_{\rm B})^2$. If the system is eclipsing and we have a measurement of the eccentricity then the lightcurve provides a model-independent {\it primary} star density ($\rho_{\rm A}$) \citep{Seager2003}. By model-independent, we mean that we derive the value solely as a function of the directly observed quantities. By combining the eclipses with radial velocities we may also derive a model-independent {\it secondary} star surface gravity ($\log g_{\rm b}$) \citep{Southworth2007}. However, we do not have enough information from direct observables to determine primary and secondary masses and radii, i.e. to break the mass-radius degeneracy \citep{Stevens2018}.

This primary star mass-radius degeneracy is comparable to that seen for exoplanet hosts \citep{Seager2003}. A typical method of breaking it is to use an evolutionary track. The Torres mass-radius relations, calibrated with eclipsing binaries, can also be used to break this degeneracy \citep{Torres2010}. Alternatively, one can derive the primary radius ($R_{\rm A}$) more empirically by using the star's distance (e.g. from Gaia \citep{GaiaCollaboration2016}) and its spectral energy distribution (SED).

In this paper we take a unique approach for the EBLM survey to try and quantify some of this model dependence for the primary star. We employ the IDL software \textsc{EXOFASTv2} which can simultaneously fit both the orbital data (photometry and radial velocities) and the models invoked to determine the primary star's parameters \citep{exofastv2}. Thus we analyse the impact of several different methods of breaking the degeneracy in the physical parameters of the larger host star. We employ two forms of evolutionary tracks:  Yonsei-Yale (YY) \citep{YY} and the Mesa Isochrone and Stellar Track (MIST) \citep{mist}. We also employ the classic \citet{Torres2010} M-R relation. Additionally, we combine each of these three methods with constraints from the SED and the Gaia DR2 parallax. By applying several methods we can estimate the systematic error being passed on to the eclipsing M-dwarf, and can determine if this might be responsible for any erroneous claims (e.g. radius inflation or outlier temperatures).

Our paper is organised into the following sections: observations (Sect.~\ref{section:observations}); methods (Sect.~\ref{section:methods}); results (Sect.~\ref{section:results}); discussion (Sect.~\ref{section:discussion}) and conclusion (Sect.~\ref{section:conclusion}).

\section{Observations}\label{section:observations}


\begin{table}
\caption{Summary of observational properties.}              
\label{Observation_table}      
\centering                                    
\begin{tabular}{c l l  l l l l l l l l l l l l l }  

\hline\hline                        
 
 & EBLM J0055$-$00 
 & EBLM J2217$-$04  \\
 
\hline 
 TIC 
 & 269504659
 & 439837578\\
EPIC 
 & 220196587
 & 206500801\\
 2MASS 
 & 00551372-0007541
 & 22175812-0451529\\\\
 
$\alpha$
& $00^{\rm h}55^{'}13.72^{"}$ 
& $22^{\rm h}17^{'}58.13^{"}$ \\
& $13.8072^{\circ}$ & $334.4922^{\circ}$ \\

$\delta$ 
& $-00^{\circ}07' 54.00^{"}$
& $-04^{\circ}51' 52.60^{"}$\\
& $-0.1317^{\circ}$ & $-4.8647^{\circ}$ \\ \\

CORALIE obs.
& 24
& 13\\

$\Delta$ t [days]
& 489
& 436 \\ \\

HARPS obs.
& 23
& 25\\

$\Delta$ t [days]
& 947
& 549\\ \\

$K2$ Campaign
 
& \#8
& \#3\\ \\

$G$-mag
& $10.9115 \pm 0.0003$
& $12.003 \pm 0.001$ \\

$V$-mag
& $10.955 \pm 0.012$
& $12.003 \pm 0.001$ \\

$G_{BP} - G_{RP}$
& 0.8342
& 1.0294 \\

Parallax [mas]
& $3.158 \pm 0.062$ 
& $2.480 \pm 0.099$ \\

\hline

\end{tabular}
\end{table} 

\begin{table}
\caption{Primary star spectroscopic parameters.}              
\label{Spectroscopic_table}      
\centering                                    
\begin{tabular}{c l l  l l l l l l l l l l l l l }  

\hline\hline                        
 
 & EBLM J0055$-$00 
 & EBLM J2217$-$04  \\
 
\hline

$\rm T_{\rm eff}$ $\rm(K)$
& $5969 \pm 85$
& $5848 \pm 85$\\

$\log g$ (dex)  
& $4.36 \pm 0.13$ 
& $4.17 \pm 0.13$ \\



Vsin$i$ (km\,s$^{-1}$)
& $ \leq 5$
& $7.97 \pm 1.35$ \\

$\rm [Fe/H]$
& $0.39 \pm 0.06$
& $0.27 \pm 0.06$\\

\hline

\end{tabular}
\end{table} 

\begin{figure*}
    \includegraphics[width=0.99\textwidth]{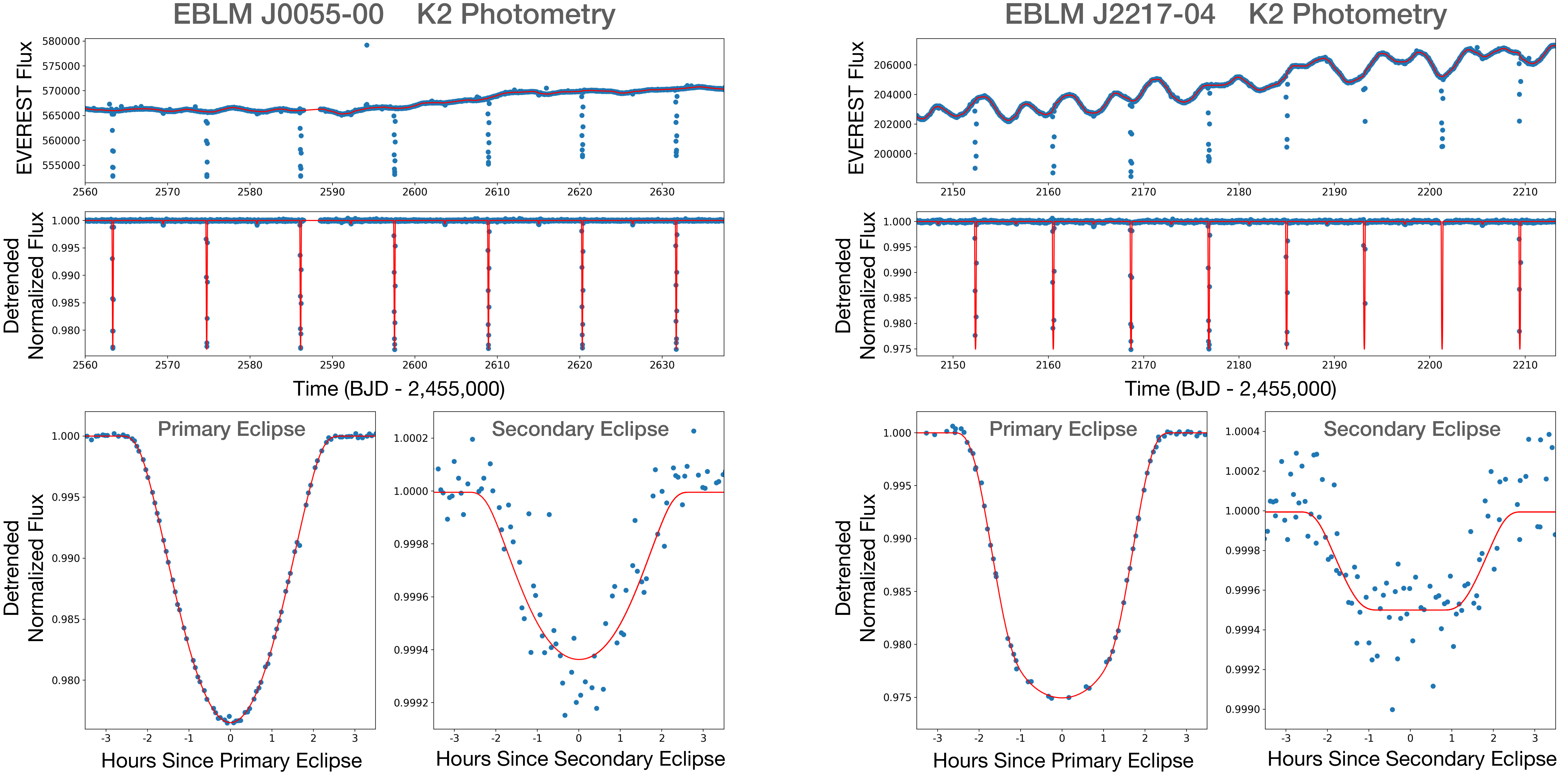}
    \caption{K2 lightcurves for EBLM J0055-00 (left) and EBLM J2217-04 (right). The top panels are the raw \textsc{EVEREST} flux in blue with the fitted red detrending line. The middle panels are the detrended light curves with the model fitted to the primary and secondary eclipses. The bottom panels show a zoomed and phase-folded primary and secondary eclipse for each binary. }\label{fig:combined_lightcurves}
\end{figure*}

\subsection{The targets}\label{observations:targets}

Our two targets are EBLM J0055-00 and EBLM J2217-04. These are both eclipsing single-lined spectroscopic binaries (SB1s) with high resolution HARPS and CORALIE spectroscopy, K2 observations, and visible primary and secondary eclipses. J2217-04 was first discovered and published in the \citet{Triaud2017} southern hemisphere EBLM catalog. J0055-00 was discovered at a similar time but was not in this catalog. Its first publication was instead in the \citet{Martin2019} BEBOP circumbinary planet search, which uses the EBLM binaries as a target list.

J0055-00A is close to a solar analog. J2217-04A is instead a bit larger and slightly evolved. In both cases the secondary star was already known to be an M-dwarf, and believed to be a fully convective star, to be confirmed in this paper. Both binaries were already known to be well-detached ($P=11.4$ days for J0055-00 and 8.2 days for J2217-04). The observational properties of these two targets are summarised in Table~\ref{Observation_table}.


\subsection{K2 photometry}\label{observations:K2}


Both targets received about 80 days of K2 photometry. We used  data products from the \textsc{EVEREST} pipeline \citep{Luger2016}. \textsc{EVEREST} is an open-source pipeline capable of producing lightcurves with precision comparable to the original Kepler mission. It uses a combination of pixel-level decorrelations to remove spacecraft pointing error and Gaussian processes to capture astrophysical variability. The \textsc{EVEREST} photometry is shown in the top panels of Fig.~\ref{fig:combined_lightcurves}.

We note that J2217-04 was observed by TESS in sector 42 (August to September 2021). This time-series is shorter and significantly less precise than the K2 data, and as such was not included in the analysis. J0055-00 has no TESS observations to date.

\begin{figure*}
    \includegraphics[width=0.45\textwidth]{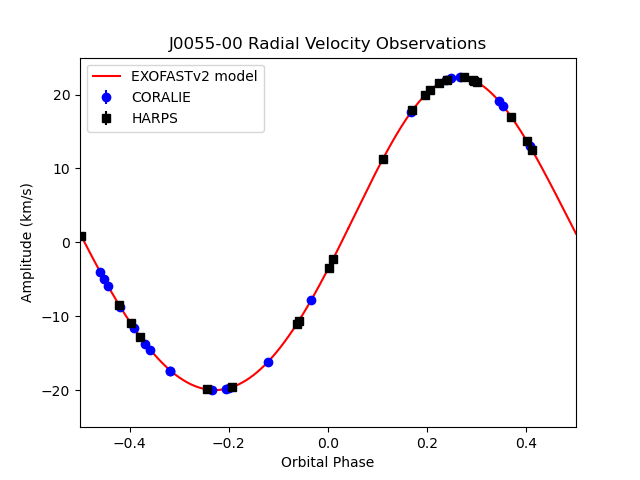}
    \includegraphics[width=0.45\textwidth]{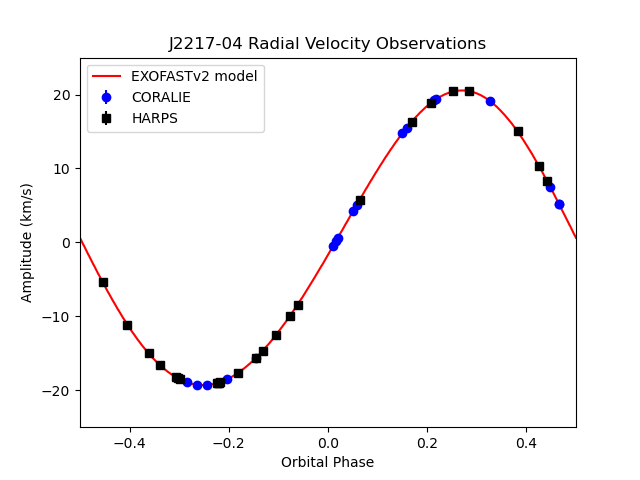}
    \caption{Radial velocity observations of both binaries, phase-folded on the fitted period. CORALIE data are shown as blue circles and HARPS data are black squares. Errorbars are on the order of m/s and are hence invisible at this scale. The red lines show the \textsc{EXOFASTv2} Model Fits. These particular fits employ the MIST evolutionary tracks for the primary star, but ultimately the RV fit is essentially independent of the primary star model.}\label{fig:RVs}
\end{figure*}

\subsection{CORALIE spectroscopy}

\coralie\ is a fiber-fed \'{e}chelle spectrograph installed on the 1.2-m Leonard Euler telescope at the ESO La Silla Observatory and has a resolving power R = 50,000\,--\,60,000 \citep{2001A&A...379..279Q,2008ApJ...675L.113W}. The spectra used in this study were all obtained with an exposure time $t_{\rm exp} = 600$\,s. The spectra for each star were processed with the \coralie\ standard reduction pipeline \citep{1996A&AS..119..373B}. Radial velocity measurements were obtained using standard cross-correlation techniques (using numerical masks) and checked for obvious outliers \citep{Triaud2017}.

\subsection{HARPS spectroscopy}

\harps\ is a fiber-fed \'{e}chelle spectrograph installed on the ESO 3.6-m telescope at the ESO La Silla Observatory and has a resolving power R = 115,000  \citep{2003Msngr.114...20M}. \harps\ spectra for J0055$-$00 and J2217$-$04 were obtained over 2 programs as part of the BEBOP search for circumbinary planets. Data reduction follows the standard HARPS pipeline, which is similar to that of CORALIE.





\subsection{Lucky imaging}\label{subsec:lucky_imaging}

The lucky-imaging technique (e.g. \citealt{2006A&A...446..739L}) was used to obtain high-resolution images of  J0055-00 and  J2217-04 in July 2017, in order to search for stars contributing contaminating light, as well as potential bound companions to the eclipsing binaries. The observations were conducted using the Two Colour Instrument (TCI) on the Danish 1.54-m Telescope at La Silla Observatory. The TCI consists of two Electron Multiplying CCDs capable of imaging simultaneously in two passbands at a frame rate of $10$\,Hz, with a $40"\times40$" field of view. The `red' arm has a passband similar to a combined $i+z$ filter or the Cousins $I$ filter, whilst the `visible' arm has a mean wavelength close to that of the Johnson $V$ filter. A detailed description of the instrument  can be found in \citet{2015A&A...574A..54S} and  the lucky imaging reduction pipeline is described by \citet{2012A&A...542A..23H}. The observations and data reduction were carried out using the method outlined in \citet{2018A26A...610A..20E}, and is briefly described here. All targets were observed for 170\,s. The raw data were reduced automatically by the instrument pipeline, which performs bias and flat frame corrections, removes cosmic rays, and determines the quality of each frame, with the end product being ten sets of stacked frames, ordered by quality. The data were run through a custom star-detection algorithm that is described in \citet{2018A26A...610A..20E}, which is designed to detect close companion stars that may not be fully resolved. 

Overall, we determined that there were no nearby stars that would contaminate the K2 photometry. This is important since such contamination would dilute the primary and secondary eclipse depths, leading to erroneous measurements of radius and effective temperature.

\section{Methods}\label{section:methods}

\subsection{Primary star spectroscopic analysis}\label{subsec:primary_star_spectroscopy}

Spectra were co-added onto a common wavelength scale to increase signal-to-noise prior to spectral analysis. Each co-added spectrum was analysed with the spectral analysis package iSpec \citep{2014A&A...569A.111B,2019MNRAS.486.2075B}. We used the synthesis method to fit individual spectral lines of the co-added spectra. We used the radiative transfer code SPECTRUM \citep{1999ascl.soft10002G} to generate model spectra with MARCS model atmospheres \citep{2008A&A...486..951G}, version 5 of the GES (GAIA ESO survey) atomic line list provided within iSpec and solar abundances from \citet{2009ARA&A..47..481A}. The H$\alpha$, Na\,I\,D and Mg\,I\,b lines were used to infer the effective temperature (\teff) and gravity (\logg) while Fe\,I and Fe\,II lines were used to determine the metallicity \feh\ and the projected rotational velocity \vsini. Trial synthetic model spectra were fit until an acceptable match to the data was found. Uncertainties were estimated by varying individual parameters until the model spectrum was no longer well-matched to the data. These values are catalogued in Table~\ref{Spectroscopic_table}. They are used as priors in the \textsc{EXOFASTv2} fits.


\subsection{Lightcurve preparation}

 We obtained the \textsc{EVEREST} lightcurves for each target from the Barbara A. Mikulski Archive for Space Telescopes (MAST).\footnote{\url{archive.stsci.edu}} We then conducted our own detrending to remove any remaining out of eclipse variability. We used the \textsc{Wotan} package \citep{Hippke2019} and a Tukey's biweight filter with a window length equal to four times the primary eclipse duration, such that the eclipse depths would be preserved. The raw and detrended lightcurves are shown in Fig. \ref{fig:combined_lightcurves}.

\subsection{Fitting with {\tt \textsc{EXOFASTv2}}}

In order to determine the physical properties of both the larger host star and the orbiting M-dwarf, we use the IDL based \textsc{EXOFASTv2} \citep{exofastv2} fitting package. \textsc{EXOFASTv2}, originally designed for exoplanet characterisation, simultaneously models both the host star and orbiting body by jointly fitting radial velocity and photometric observations using a differential evolution Markov Chain Monte Carlo (MCMC) method \citep{exofastv2}. \textsc{EXOFASTv2} employs the \citet{Agol2020} eclipse models and uses the Claret tables at each step to fit the quadratic limb darkening coefficients $h_{1}$ and $h_{2}$ \citep{claret2017,claret2011}.  \textsc{EXOFASTv2} is able to use several different methods of breaking the degeneracy between the mass and radius of the host star and constraining its properties to provide physical solutions. In this paper we use three constraints: the Torres \citep{Torres2010} relations, the YY Evolutionary Tracks, and the MIST Evolutionary Tracks. In all three cases we do a combined fit with the spectral energy distribution (SED) and Gaia parallax. Each of the mass-radius breaking constraints are only used to model the primary star. The orbiting M-dwarf is fit using the transit and radial velocity observations.

\subsubsection{Spectral energy distribution plus parallax}\label{subsubsec:SED}

\begin{figure}
    \includegraphics[width=0.40\textwidth]{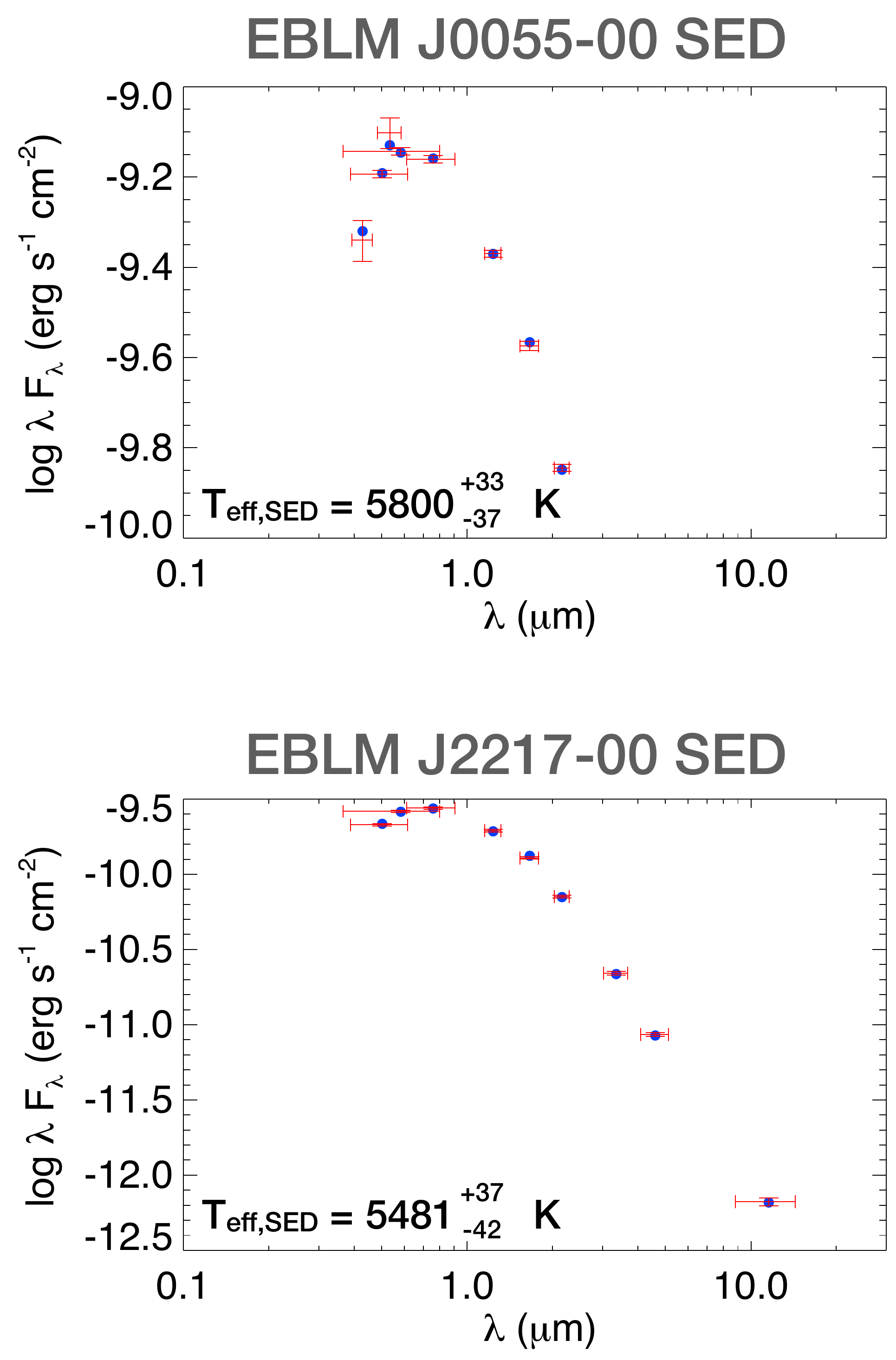}
    \caption{Spectral energy distributions for both targets (red data points), as fitted by \textsc{EXOFASTV2} (blue model points). For J0055-00 the SED temperature is 169 K cooler than that from spectral fitting, which is a $\approx 2\sigma$ difference based on the spectroscopy uncertainty. For J2217-04 the SED temperature is 367 K cooler, which is a $\approx 4\sigma$ difference. We suspect that reddening may affect the SED-derived temperatures. Ultimately the temperature derived by \textsc{EXOFASTv2} combines the spectroscopic prior with fits to the SED, evolutionary tracks, K2 photometry and radial velocity data. This temperature is typically in between the SED and spectroscopy values.  }\label{fig:seds}
\end{figure}

\textsc{EXOFASTv2} can use broad band photometry to model the SED of a star which provides a measure of the star's bolometric flux \citep{exofastv2}. The SED is mostly independent of stellar models \citep{Stassun_2018}. This means it can be used in conjunction with the following models without double counting information about the star. We add the 0.082 mas systematic parallax offset found in the Gaia DR2 inferred by \citep{Stassun_2018}. For J2217-04 we include magnitude observations from GAIA DR2 \citep{gaia2018}, 2MASS \citep{cutri_2003}, and WISE \citep{cutri_2014}. J0055-00 uses the same suite of observations plus additional magnitudes from Tycho \citep{hoeg_2000}. The SED depends only weakly on the surface gravity $\log g$ and metallicity of the star, but does place important constraints on \teff\ and $V$-band extinction $A_V$. By fitting the SED and inferring $A_V$, we can infer the bolometric flux, which when combined with \teff, yields  $(R/d)^2$ \citep{Stassun2017}.  The stellar radius of the primary star, $R_A$, can then be inferred from the distance derived by the Gaia parallax. Our two SED fits are shown in Fig.~\ref{fig:seds}.
  
\subsubsection {Torres semi-empirical relations}

The Torres relations are a semi-empirical framework based on the work of \cite{Torres2010}. The study looked at 95 detached double-lined eclipsing binary systems (and the $\alpha$ Centauri system) and found empirical relationships between the mass and radii of these stars and $\log{g_*}$, \teff, and [Fe/H]. Double-lined eclipsing binary systems allow for accurate measurements of the mass and radii of both stars, with the need for an external constraint on the distance. (Indeed, it is possible to infer the distances to these systems from the fit to the eclipses and radial velocities.) 
The Torres relations apply primarily to unevolved or somewhat evolved main-sequences stars. They do not apply to giants or pre main-sequence stars. Furthermore, only a handful of the stars in the \citet{Torres2010} sample were low-mass M stars, and thus one should be wary of applying them to low-mass stars. To be clear, in our study we apply the Torres relations to the {\it primary} star only.

The Torres relations yield masses and radii that are accurate to 6\% and 3\%, respectively, based on the scatter of the measured values of these quantities relative to those predicted by the relations. With measurements of $\log g$, \teff, and [Fe/H], these relations can be used to quickly estimate $M$ and $R$.


\subsubsection{ The MIST evolutionary tracks} 

\begin{figure}
    \includegraphics[width=0.49\textwidth]{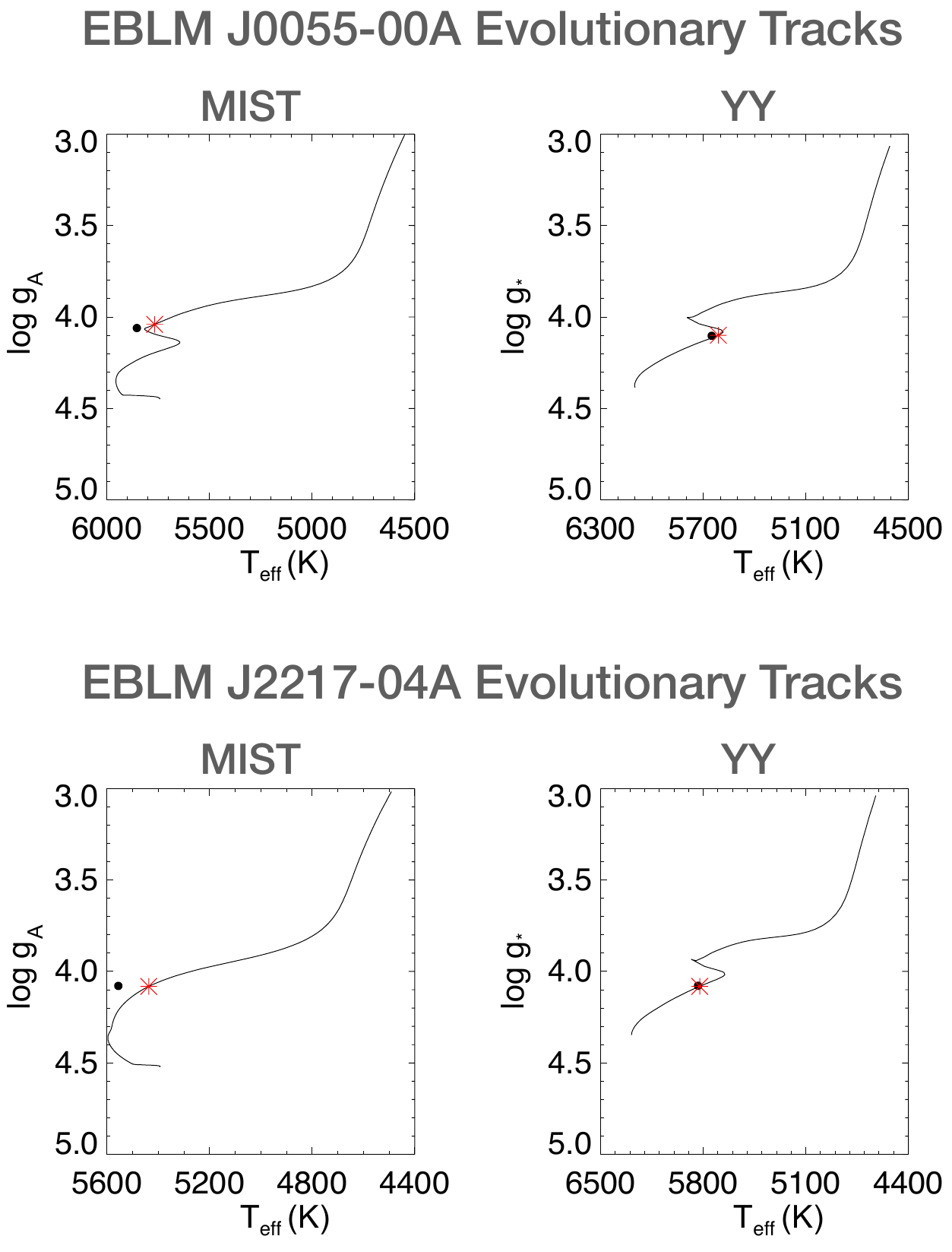}
    \caption{The stellar evolutionary tracks for both J0055-00A (top panels) and J2217-04A (lower panels). The black lines represent the evolutionary track for the best fit stellar mass. The black dot is  $\rm \log{g_{\rm A}}$ and \teff\ as fit by \textsc{EXOFASTv2}. The red asterisk represents the value predicted purely by the evolutionary track. The position of both targets on the Hertzsprung-Russell diagram near an inflection point on the evolution track leads to bimodality in the stellar age, which propagates into a bimodality into $M_{\rm A}$ and $R_{\rm B}$, and in turn $M_{\rm B}$ and $R_{\rm B}$.}   \label{fig:evolutionary_tracks}
\end{figure}

The MESA Isochrone and Stellar Tracks (MIST) models \citep{mist} are the suggested default for \textsc{EXOFASTv2} \citep{exofastv2}. These models are valid for stars between $0.1~M_\odot$ to $300~M_\odot$, starting at 100,000 years in age thus including pre main-sequence stars \cite{mist}. The MIST models are used to estimate the properties of the primary star in this work. The stellar evolutionary tracks are computed using a grid of initial mass, initial [Fe/H], and evolutionary phase \citep{exofastv2}. Fits of MIST evolutionary tracks to our two targets are shown in the left-most panels in Fig.~\ref{fig:evolutionary_tracks}. 

\subsubsection{Yonsei-Yale evolutionary tracks} 
The Yonsei-Yale (YY) stellar evolutionary tracks predict the evolution of stars from pre main-sequence to predict the luminosity, color, \teff, and radius as a function of mass, age, and metallicity of the star \citep{YY}. The YY tracks are used here to estimate the properties of the primary star in our binaries. Fits of YY evolutionary tracks to our primary stars in our two target systems are shown in the right-most panels in Fig.~\ref{fig:evolutionary_tracks}.

\subsubsection{Applying \textsc{EXOFASTv2}} 


For each model (MIST, YY and Torres) we begin by creating a fit solely for the primary star (i.e. solely fitting the SED and either the evolutionary track or Torres relation). We include spectroscopically-derived priors on effective temperature and [Fe/H] from Table~\ref{Spectroscopic_table}. We then use this converged primary-only fit as a starting point for a second \textsc{EXOFASTv2} model, where we fit both stars by incorporating the K2 lightcurve and the radial velocity observations.


We run the system fit until either it converges to a Gelman-Rubin score of $<1.05$ or we reach a maximum number of 15,000 steps. The Gelman-Rubin statistic examines the variance within individuals chains and the variance between groups of chains. When the variances between individual chains and groups of chains are similar the simulation is considered well mixed. Large differences in these variances are considered non-converged. Values of the Gelman-Rubin statistic close to one are often considered converged. Therefore, we adopt a convergence threshold of 1.05. If the convergence criterion is not achieved on the first iteration, we use the results of the first iteration as starting values, without associated uncertainties, in a second iteration, as recommended by \citet{exofastv2}. We repeat this process until convergence is achieved


\subsection{Determining the M-dwarf temperatures}

\begin{figure*}
    \includegraphics[width=0.99\textwidth]{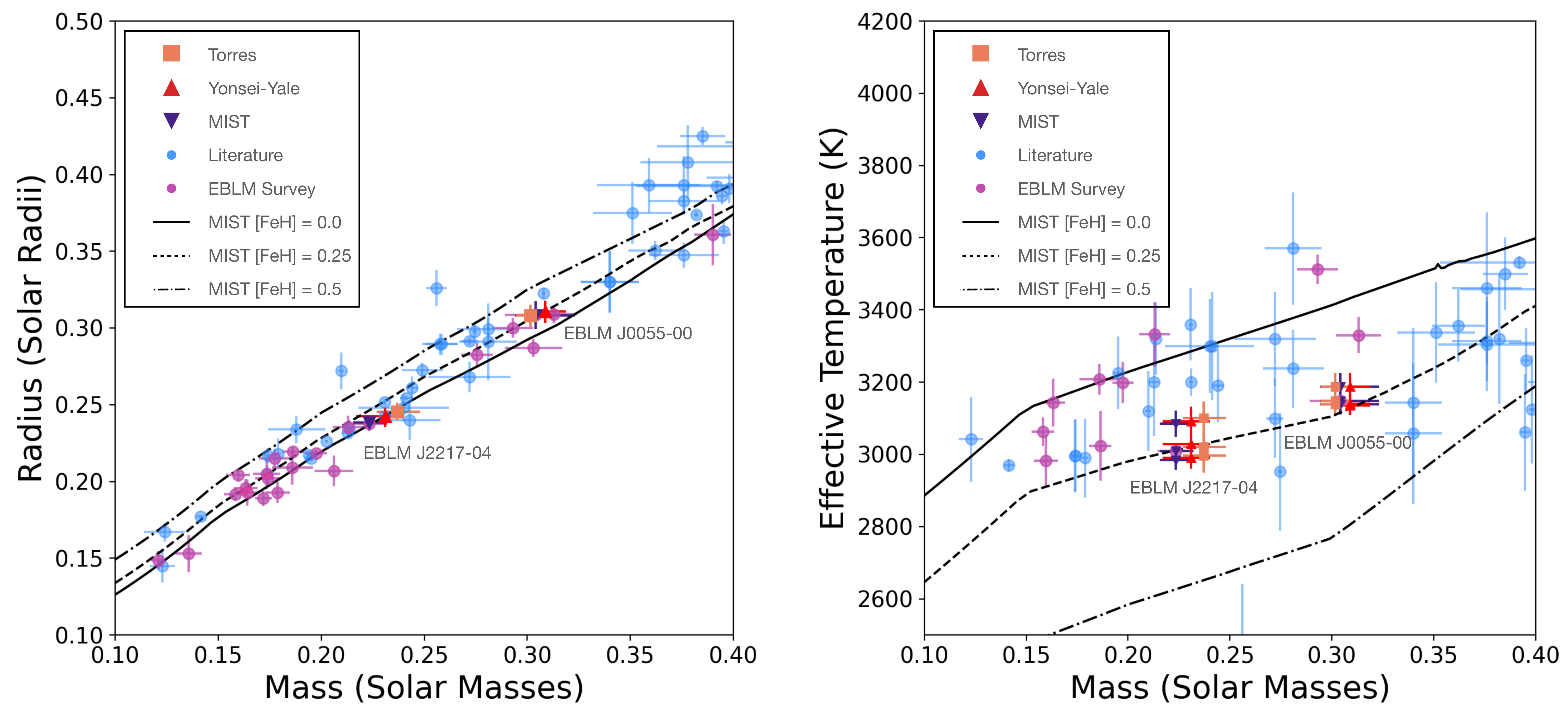}
    \caption{Fundamental stellar parameters for J0055-00 and J2217-04 using three different models (MIST, YY and Torres), and a comparison with literature M-dwarfs. The literature stars (blue) are compiled from \citet{Parsons2018},  with the addition of published stars from our EBLM survey (purple). Only stars with masses and radii better than 10\% and measured temperatures are included. On both plots the black curves indicate MIST stellar models for [Fe/H]=0.0, [Fe/H]=0.25 and  [Fe/H]=0.5. On the left is mass-radius and on the right is mass-effective temperature. We see no sign of radius inflation, regardless of the model used. For the mass-temperature plot we show three temperatures for each model, where the difference is the choice of $T_{\rm eff,A}$ used in Eq.~\ref{eq:sec_eclipse_depth2}. The hottest $T_{\rm eff,B}$ comes from the spectroscopic $T_{\rm eff,A}$ (Table~\ref{Spectroscopic_table}. The coldest comes from the SED-only fit (Fig.~\ref{fig:seds}. The central value is our nominal value, coming from the joint \textsc{EXOFASTv2} fit. Both targets are above-Solar metallicity, in line with their low temperature at a given mass. Changing models has minimal effect on the M-dwarf temperature but changing the primary star's temperature has an appreciable effect.}   \label{fig:mass_radius_temperature}
\end{figure*}

\textsc{EXOFASTv2} does not explicitly fit the secondary temperature. We derive $T_{\rm eff,B}$ from the secondary eclipse depth $D_{\rm sec}$ using a method applied in other studies such as \citet{Swayne2020,Swayne2021}. We invoke the following relationship between eclipse and depth and effective temperature:

\begin{align}
    \label{eq:sec_eclipse_depth}
    D_{\rm sec} &=k^2S + A_{\rm g} \left(\frac{R_{\rm B}}{a}\right)^2, \\
    \label{eq:sec_eclipse_depth2}
    &= \left(\frac{R_{\rm B}}{R_{\rm A}}\right)^2\frac{\int \tau(\lambda)F_{\rm B,\nu}(\lambda,T_{\rm eff,B})\lambda d\lambda}{\int \tau(\lambda)F_{\rm A,\nu}(\lambda,T_{\rm eff,A})\lambda d\lambda} + A_{\rm g} \left(\frac{R_{\rm B}}{a}\right)^2,
\end{align}
where $k=R_{\rm B}/R_{\rm A}$ is the radius ratio, $S$ is the surface brightness ratio, $A_{\rm g}$ is the geometric albedo, $\tau(\lambda)$\footnote{\url{http://svo2.cab.inta-csic.es/svo/theory/fps3/index.php?id=Kepler/Kepler.K}} is the Kepler transmission function as a function of wavelength $\lambda$ and $F$ is the normalised flux. Equation ~\ref{eq:sec_eclipse_depth2} has been used in past studies such as \citet{Charbonneau2005,Canas2022}. The transmission functions are set up for the photon-counting instrumental CCDs. To correctly gain the instrumental flux rather than the number of photons we introduce a factor of $\lambda$ as set-out in Appendix A2 of \citet{Bessell2012}.

The secondary eclipse depth has two contributions: the brightness (and hence temperature) of the secondary, and its reflectivity. However, we ignore any reflective light (i.e. $A_{\rm g}(R_{\rm B}/a)^2\approx 0$) since with $A_{\rm g}\sim0.1$ \citep{Mazeh2010,Cowan2011,Canas2022} the effect is roughly 15 ppm, whereas $D_{\rm sec}\approx500-600$ ppm.

For a total secondary eclipse, i.e. the secondary star is completely occulted, the secondary eclipse depth will correspond to the flux ratio of the two stars. This is the case for J2217-04 (flat-bottomed eclipse in Fig.~\ref{fig:combined_lightcurves}). For J0055-00 the secondary eclipse is grazing, as seen in Fig.~\ref{fig:combined_lightcurves} where it has a U-shaped secondary eclipse. \textsc{EXOFASTv2} accounts for grazing secondary eclipses so its outputted value of $D_{\rm sec}$ is equal to what it would be for a total secondary eclipse. The J0055-00 secondary eclipse is only slightly grazing; we calculate that 90\% of the star is occulted. Using the code \textsc{ellc} we test if limb darkening of the M-dwarf affects the measured $D_{\sec}$. We determine that limb darkening only has a 16 ppm effect, which is less than the $\approx 32$ ppm measurement precision, so we can safely ignore it.


Based on our lucky imaging work (Sect.~\ref{subsec:lucky_imaging}) and the small size of the Kepler pixels (4") we do not apply any dilution corrections (see Martin et al. under rev. for details of when such a process is necessary).

To determine the integrated fluxes in Eq.~\ref{eq:sec_eclipse_depth2} we first take the  \textsc{EXOFASTv2} outputted values for $T_{\rm eff,A}$, $\log g_{\rm A}$ and [Fe/H] and create a high resolution PHOENIX model spectrum for the primary star. This spectrum is then convolved with the Kepler transmission function. We integrate the spectrum to determine the brightness of the primary star within the Kepler bandpass. We then calculate the brightness of the secondary star in the same way, using a grid of secondary star temperatures between 2300 and 4000 K. We use a  Brent minimisation routine to solve for $T_{\rm eff,B}$. This process is repeated over a distribution of the observables with $1\sigma$ errors so we derive a posterior distribution for $T_{\rm eff,B}$. 


\section{Results}\label{section:results}

Our results for J0055-00 and J2217-04 are tabulated in Tables~\ref{tab:0055params} and ~\ref{tab:2217params}, respectively. Each table shows results for all three model choices for the primary star (MIST, YY and Torres). Even though our \textsc{EXOFASTv2} models simultaneously fit stellar parameters, photometry and the radial velocities, for clarity we separate the parameters into different categories. 

We show the fits to the photometry, radial velocities and stellar evolutionary tracks in Fig.~\ref{fig:combined_lightcurves}, ~\ref{fig:RVs} and ~\ref{fig:evolutionary_tracks}, respectively.


We illustrate the differences in the  parameters between models in Figs.~\ref{fig:J0055exofastparam} and ~\ref{fig:J2217exofastparam}. It is seen that the primary star mass and radius vary quite significantly between models. This imprints a similar variation on the parameters we are most interested in: the secondary star (M-dwarf) mass and radius. Contrastingly, the orbital parameters are essentially model-independent. The differences between the different models are discussed in Sect.~\ref{subsec:discussion_differentmodels} and tabulated in Table~\ref{table:model_differences}.

\section{Discussion}\label{section:discussion}

\subsection{Literature comparison}

In Fig.~\ref{fig:mass_radius_temperature} we plot our two targets against known M-dwarfs with precise stellar parameters from the literature. We base the comparison on a literature compilation in \citet{Chaturvedi2018}, although we exclude three outlier temperatures since a concurrent study finds those values to be erroneous (Martin et al., under rev.). For context, we show theoretical MIST stellar models at both solar ([Fe/H]=0) and above-Solar ([Fe/H]=0.25) metallicities. Note that whilst we only directly measure the metallicity of the primary (through spectroscopic fitting), it is assumed that an eclipsing binary will have formed at the same time from the same material, and hence the secondary star (M-dwarf) will have the same metallicity. Both targets have a constrained above-solar metallicity from spectroscopy. Owing to the precise K2 spectroscopy, the radii of our two targets are two of the best-characterised in the EBLM program and indeed the literature.


\subsection{Differences in primary star models}\label{subsec:discussion_differentmodels}

\subsubsection{Absolute M-Dwarf mass and radius}\label{subsubsec:massradius}


We investigate the impact of our choice of model used to characterise the primary star. Since we do not know a priori which model is ``best'', we dub this the ``model uncertainty'' $\delta_{\rm model}$.  We define this simply to be the difference between the maximum and minimum value for a given parameter. We then compare this with the mean ``fitting uncertainty'' $\langle\delta_{\rm fit}\rangle$, which is calculated as the average $1\sigma$ error bar calculated across the three different models. Since the posterior distributions about the median are asymmetric for many of the parameters, we average the amplitude of the positive and negative error. In Table~\ref{table:model_differences} we quantify the fitting and model uncertainties in the fundamental M-dwarf parameters.

For J0055-00 the largest model uncertainty is in the mass, making a 2.4\% contribution, but it is still smaller than the 3.2\% fitting uncertainty. For the radius the model uncertainty, if added in quadrature, would make a negligible contribution.

In J2217-04 the fitting uncertainties are similar to J0055-00, but the model uncertainties are significantly higher. In fact, for both mass and radius the model uncertainty is larger than the fitting uncertainty. This may be a consequence of significantly different evolutionary track fits between MIST and YY, as demonstrated in  Figure~\ref{fig:evolutionary_tracks}.



In our study of these two targets we did not consistently find one of our three methods of primary star characterisation to consistently produce larger or smaller masses and radii for the M dwarf and primary star. In the case of J0055-00 the Torres and SED produced smallest estimates for M dwarf mass and radius while the YY and SED method produced the largest estimates. Conversely in the case of J2217-04 we find that the Torres and SED approach produced the largest estimates for M dwarf mass and radius while the MIST and SED approach produced the smallest estimates. A larger sample would be required to determine if there are systematic biases between the models.
 
The impact of model selection is seen not only in the median value and the uncertainties in the fit parameters, but also in the shape of the posterior. For the primary star's mass and radius the MIST and YY fits are bimodal. The fit using the Torres relations is unimodal. This bimodality in the MIST and YY fits is consequently imparted on the secondary star's mass, although not its radius.  The root of the bimodality in $M$ and $R$ can be found in the stellar evolutionary track fitting, as demonstrated in Fig.~\ref{fig:evolutionary_tracks}. For J0055-00 the primary star is more massive than the Sun ($\approx 1.3M_\odot$). As stars this mass evolve off the main-sequence there is a ``hook'' in its evolution on the HR diagram before moving to the sub-giant branch. The exact shape of this hook will be a function of how core overshooting is treated \citep{Woo2001,Kippenhahn2012}, and this treatment will likely be different between MIST and YY models. Indeed, the MIST and YY models are unable to exactly pinpoint the star's evolutionary stage; MIST has the star above the hook and YY has it below. This uncertainty leads to a bimodality in the derived age, which imparts the bimodality in the mass and radius. For J2217-04 the star is closer to Solar mass. The MIST model fits a lower mass star ($1.065M_\odot$), for which there is no such hook, leading to a more unimodal distribution in mass and radius, but one smaller than for the YY model. Since the Torres models do not come from evolutionary track fitting, their results will always be unimodal. Some of these effects were seen in an earlier EBLM study by \citet{Gill2019}.

\begin{figure*}
    \includegraphics[width=15cm]{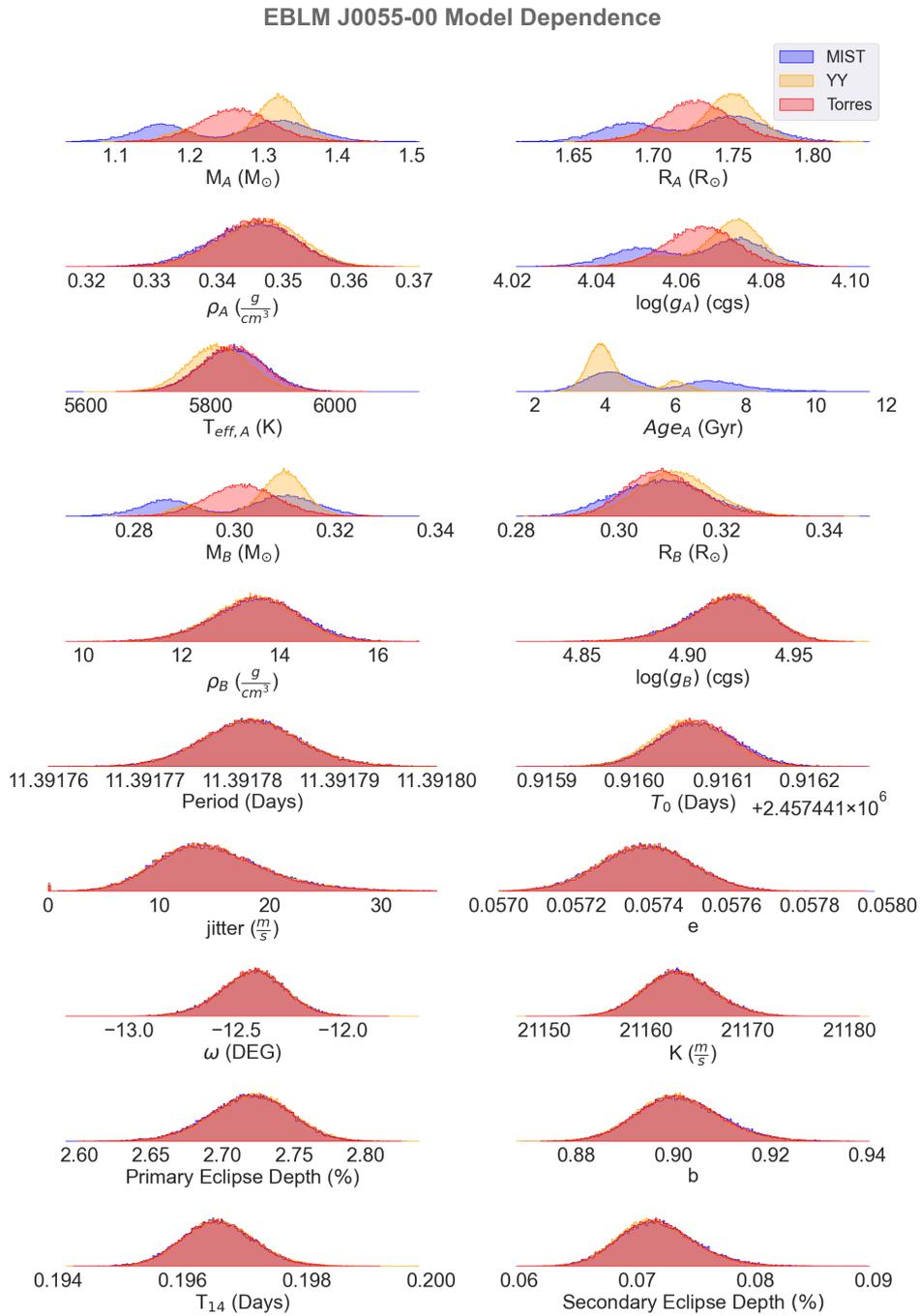}
    \caption{Model-dependence of the fundamental fitted parameters of EBLM J0055-00 for the primary star (A), the M-dwarf secondary star (B) and the binary orbital parameters. We see that uncertainty in the model choice can significantly change the primary star mass, radius and surface gravity. The primary star density, however, is derived model-independently. The directly observed orbital parameters (e.g. $K$ and eclipse depths) are also model-independent.}\label{fig:J0055exofastparam}
\end{figure*}

\begin{figure*}
    \includegraphics[width=15cm]{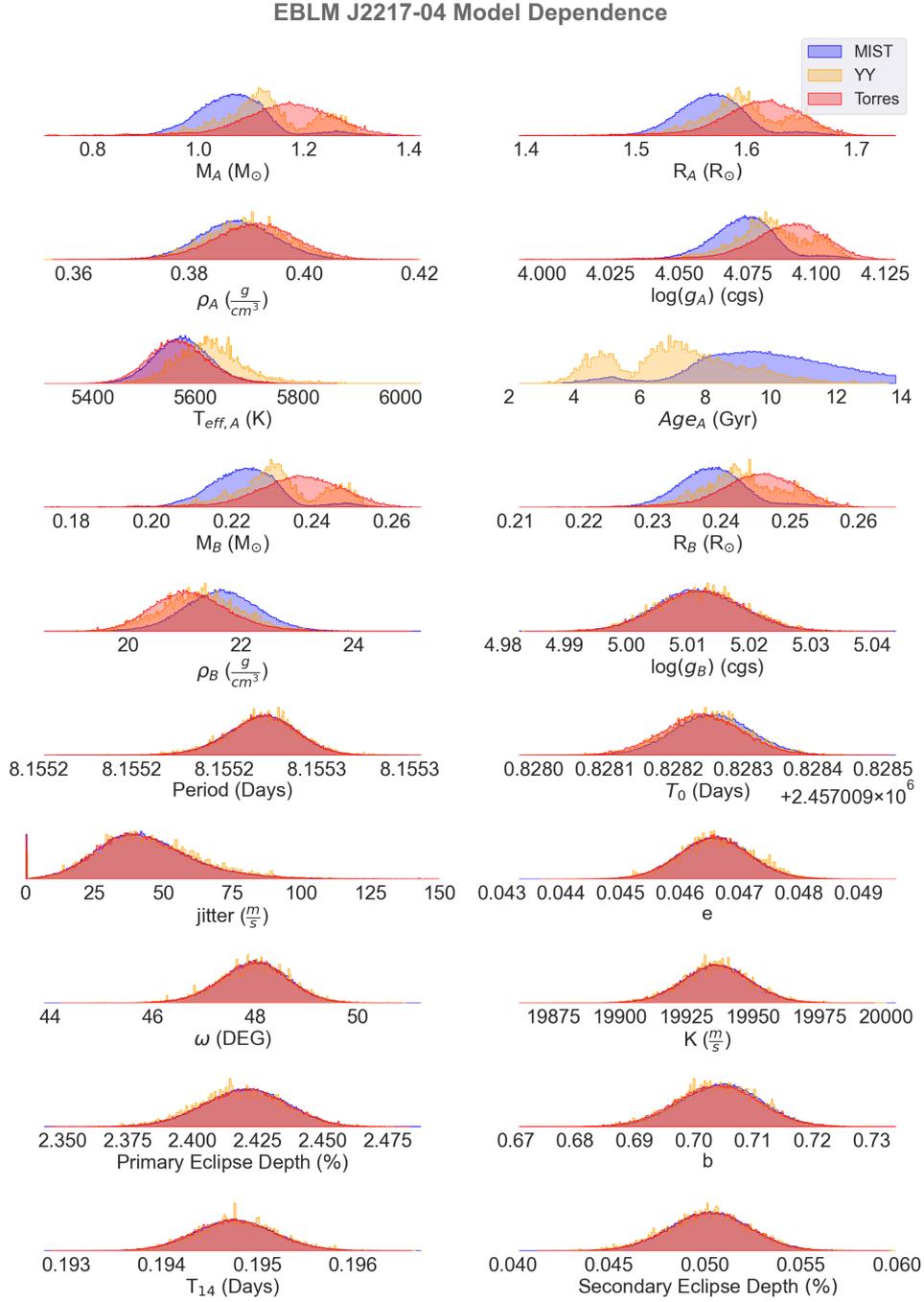}
    \caption{Same as Fig.~\ref{fig:J0055exofastparam} but for J2217-04.}\label{fig:J2217exofastparam}
\end{figure*}

Finally, we note that changing the primary star characterisation method affects the stellar parameters but the direct observables (e.g. period, eclipse depths, radial velocity amplitude) and derived orbital parameters (e.g. $e$ and $\omega$) are largely insulated from the effect of using different methods to break the degeneracy in the primary star mass and radius. This is why in Figs.~\ref{fig:J0055exofastparam} and \ref{fig:J2217exofastparam} the histograms in the lower panels are typically directly overlapping. Finally, as expected from \citet{Seager2003,Southworth2007}, the primary star density ($\rho_{\rm A}$) and the secondary star surface gravity ($\log g_{\rm B}$) are model-independent.

\subsubsection{Radius ``inflation''}\label{subsubsec:massradius}

We find that uncertainty in model choice induces significant model uncertainties in $M_{\rm B}$ and $R_{\rm B}$, which are  comparable to the fitting uncertainty. In Fig. \ref{fig:mass_radius_temperature} (left) we show the mass and radius for all three models for both targets compared to values in the literature and MIST evolutionary models. Different models tend to move the data point diagonally, i.e. parallel to the theoretical mass-radius relation. This has a significant consequence that the choice of model for the primary star will not make the secondary star seem more or less inflated. Otherwise put, poor model selection likely cannot be the culprit for the phenomenon of radius ``inflation''.

\subsubsection{M-Dwarf effective temperature}\label{subsubsec:massradius}

The M-dwarf effective temperatures are the most precisely fitted fundamental parameter, with a fitting uncertainty of $\approx1\%$ for both targets. This makes J0055-00B and J2217-04B two of the best characterised M-dwarf temperatures. The model uncertainties are even smaller (Table~\ref{table:model_differences}). It makes sense that the model uncertainty in $T_{\rm eff,\rm B}$ is less of a factor than for $M_{\rm B}$ and $R_{\rm B}$ because it is largely derived as a function of direct observables $D_{\rm sec}$ and $k=R_{\rm B}/R_{\rm A}$. The model dependence mainly comes from the fitted primary effective temperature, although even then there is not a significant difference in $T_{\rm eff,A}$ between the three models.

The biggest difference in $T_{\rm eff,A}$ is between the value derived from spectroscopy (Sect.~\ref{subsec:primary_star_spectroscopy}) and that  from the SED (Sect.~\ref{subsubsec:SED}). They are discrepant by a few 100 K, which corresponds to several $\sigma$. For both systems the spectroscopy values are hotter, possibly due to reddening effects on the SED. Our nominal values of $T_{\rm eff,A}$ (as in Tables~\ref{tab:0055params} and \ref{tab:2217params}) come from a combined \textsc{EXOFASTv2} fit to the SED, evolutionary tracks, K2 photometry and radial velocities, using the spectroscopy values as a prior. Consequently, these temperatures are in between the spectroscopy-only and SED-only values.

We test the impact of different assumptions on $T_{\rm eff,A}$ as another type of model uncertainty. For all three models (MIST, YY and Torres) we re-calculate $T_{\rm eff,B}$ using values of $T_{\rm eff,A}$ from the relatively hot spectroscopy fit and the relatively cold SED fit. These values are plotted in Fig.~\ref{fig:mass_radius_temperature}. It is shown that differences in the assumed primary star temperature induce a bigger uncertainty than the choice of primary star model. It is also bigger than the fitting uncertainty.

There have been several M-dwarfs published with temperatures that are 500-1000 K hotter or colder than expected from both models and the majority of the literature. \citet{MaqueoChew2014}'s outlier result for EBLM J0113+31B was later corrected by \citet{Swayne2020}. Outliers for KIC 1571511B \citep{Ofir2012} and HD 24465B \citep{Chaturvedi2018} have been recently corrected by Martin et al. (under rev.). In neither \citet{Swayne2020} or Martin et al. (under rev.) was the culprit of the erroneous result identified definitively.

In our paper the model uncertainties on $T_{\rm eff,B}$ are 8 K and 19 K for the two targets, and hence could not contribute to such discrepant results. The uncertainties related to the choice of $T_{\rm eff,A}$ are larger, 50 K and 101 K, but still too small to explain the outliers. As a final test, we re-fitted $T_{\rm eff,B}$ under the assumption of [Fe/H]$=0$ in our \textsc{PHOENIX} models. We obtain results that differ by only 10's of Kelvin. A metallicity uncertainty therefore also cannot explain those literature outliers.

\renewcommand{\arraystretch}{1.5}
\begin{table}
\caption{Mean fundamental parameters, mean fitting uncertainties and model uncertainties.}  \label{table:model_differences}
\begin{tabular}{llll}
\hline
Value & Unit & EBLM J0055-00   & EBLM J2217-04  \\
                                                    \hline \hline
$\langle M_{\rm B} \rangle $ &$(M_\odot)$                  & 0.3048          & 0.230          \\
$\langle \delta_{\rm fit}M_{\rm B} \rangle$ &$(M_\odot)$  & 0.0097 (3.2\%)  & 0.010 (4.3\%)  \\
$\delta_{\rm model} M_{\rm B}$ &$(M_\odot)$          & 0.0073 (2.4\%)  & 0.014 (6.1\%)  \\
\hline
$\langle R_{\rm B}\rangle$ &$(R_\odot)$                  & 0.3094          & 0.2424         \\
$\langle \delta_{\rm fit} R_{\rm B}\rangle$ &$(R_\odot)$ & 0.0078 (2.52\%) & 0.0055 (2.3\%) \\
$\delta_{\rm model} R_{\rm B}$ &$(R_\odot)$          & 0.0025 (0.81\%) & 0.0069 (2.8\%) \\
\hline
$\langle \delta T_{\rm eff,B} \rangle$ &(K)               & 3145            & 3019           \\
$\langle \delta_{\rm fit} T_{\rm eff,B} \rangle$ &(K)     & 31 (0.99\%)     & 32 (1.1\%)     \\
$\delta_{\rm model}T_{\rm eff,B}$& (K)               & 8 (0.25\%)      & 19 (0.63\%)   \\
$\delta_{\rm primary~temp}T_{\rm eff,B}$& (K)               & 50 (1.59\%)      & 101 (3.35\%)   \\
\hline
\end{tabular}
\end{table}

\section{Conclusion}\label{section:conclusion}

We present a detailed analysis of two eclipsing binaries observed by K2. Both stars contain a G primary and a fully convective M-dwarf secondary. Given the exquisite K2 photometry of both primary and secondary eclipses, combined with high resolution CORALIE and HARPS radial velocities, we derive some of the most precise M-dwarf fundamental parameters in the literature. The fitted errors for the M-dwarf's mass, radius and effective temperature are on the order of $5\%$, $3\%$ and $1\%$, respectively. These two targets have arguably the best photometry of any in the EBLM sample, leading to two of our most precisely measured radii and temperatures.

Both targets are compatible with theoretical mass-radius models, and hence do not show signs of the infamous ``radius inflation''. The effective temperatures are a little colder than most of the literature but match theoretical expectations for above-solar metallicity.

For the first time in the EBLM survey we test different models for the breaking the mass-radius degeneracy for the primary star: MIST evolutionary tracks, Yonsei-Yale evolutionary tracks and the \citet{Torres2010} mass-radius relationship. We determine that the choice of model can introduce model uncertainties of a few per cent. For ultra-precise M-dwarf stellar characterisation this is not negligible, and indeed can even be greater than the fitting uncertainty.

Inconsistent models in the literature, plus an underestimate of the model uncertainty, may affect our interpretations of the M-dwarf fundamental parameters. However, we argue that the choice of primary star model is unlikely to affect whether or not an M-dwarf's radius is measured as inflated.

\section*{Affiliations}
$^{1}$ Department of Astronomy, The Ohio State University, 4055 McPherson Laboratory, Columbus, OH 43210, USA \\ 
$^{2}$NASA Sagan Fellow\\
$^{3}$ Department of Physics, University of Warwick, Gibbet Hill Road, Coventry CV4 7AL, United Kingdom \\
$^{4}$ Astrophysics Group, Keele University, Staffordshire, ST5 5BG, United Kingdom \\
$^{5}$ School of Physics \& Astronomy, University of Birmingham, Edgbaston, Birmimgham, B15 2TT, UK \\
$^{6}$Department of Physical Sciences, Indian Institute of Science Education and Research, Berhampur, Odisha 760010, India \\
$^{7}$ Centre for Exoplanet Science, SUPA School of Physics and Astronomy, University of St Andrews,  North Haugh, St Andrews KY16 9SS, UK \\
$^{8}$ Astrobiology Research Unit, Université de Liège, Allée du 6 Août 19C, B-4000 Liège, Belgium \\
$^{9}$ Lowell Observatory, 1400 W. Mars Hill Rd., Flagstaff, AZ 86001, USA 
$^{10}$ Observatoire Astronomique de l'Université de Genève, Chemin Pegasi 51, Versoix, Switzerland \\
$^{11}$ Aix Marseille Univ, CNRS, CNES, LAM, 38 rue Frédéric Joliot-Curie, 13388 Marseille, France \\

\section*{Acknowledgements}

Support for AD was provided by NASA TESS Guest Investigator Programs for Cycle 4 (G04157) and Cycle 2(G022253). Support for DVM was provided by NASA through the NASA Hubble Fellowship grant HF2-51464 awarded by the Space Telescope Science Institute, which is operated by the Association of Universities for Research in Astronomy, Inc., for NASA, under contract NAS5-26555. This research is also supported work funded from the European Research Council (ERC) the European Union’s Horizon 2020 research and innovation programme (grant agreement n◦803193/BEBOP).  Partial support for AD, RRM, and BSG
was provided by the Thomas Jefferson Chair Endowment for Discovery and Space Exploration. 
MIS acknowledges support from STFC grant number ST/T506175/1.

\section*{Data Availability Statement}

All radial velocities and light curves will be made available online.

\bibliographystyle{mnras}
\bibliography{EBLMIX} 


\renewcommand{\arraystretch}{1.5}

\begin{table*}
	\centering
	\caption{Fitted parameters for the eclipsing single-lined spectroscopic binary EBLM J0055-00 using three different models for the primary star.}
	\label{tab:0055params}
	\begin{tabular}{lccc}
	\hline
	& MIST & YY & Torres	 \\
	\hline
	\hline
	{\bf Primary Star Parameters} &&&\\
	$M_{\rm A} (M_{\odot})$& $ 1.281^{+0.07}_{-0.13} $& $ 1.313^{+0.032}_{-0.068} $& $ 1.263^{+0.049}_{-0.047} $	\\
	$R_{\rm A} (R_{\odot})$& $ 1.731^{+0.033}_{-0.052} $& $ 1.746^{+0.018}_{-0.028} $& $ 1.727^{+0.023}_{-0.022} $	\\
	$\rho_{\rm A}$ (cgs)& $ 0.3455^{+0.0063}_{-0.0068} $& $ 0.3468^{+0.0058}_{-0.0062} $& $ 0.3456^{+0.0058}_{-0.0062} $	\\
	$\log g_{\rm A}$ (cgs)& $ 4.066^{+0.011}_{-0.019} $& $ 4.0712^{+0.0065}_{-0.011} $& $ 4.0647^{+0.0082}_{-0.0084} $	\\
	$T_{\rm eff,A}$ (K)& $ 5835.0^{+52.0}_{-52.0} $& $ 5812.0^{+50.0}_{-48.0} $& $ 5834.0^{+49.0}_{-50.0} $	\\
	$\rm [Fe/H]$& $ 0.386^{+0.056}_{-0.058} $& $ 0.38^{+0.061}_{-0.062} $& $ 0.388^{+0.059}_{-0.059} $	\\
	{\bf Secondary Star (M-Dwarf) Parameters} &&&\\
	$M_{\rm B} (M_{\odot})$& $ 0.304^{+0.01}_{-0.019} $& $ 0.3088^{+0.0047}_{-0.01} $& $ 0.3015^{+0.0072}_{-0.007} $	\\
	$R_{\rm B} (R_{\odot})$& $ 0.3085^{+0.0090}_{-0.0092} $& $ 0.3110^{+0.0075}_{-0.0070} $& $ 0.3086^{+0.0077}_{-0.0068} $	\\
	$\rho_{\rm B}$ (cgs) &$ 13.55^{+0.87}_{-0.91} $& $ 13.47^{+0.81}_{-0.87} $& $ 13.53^{+0.84}_{-0.88} $	\\
	$\log g_{\rm B}$ (cgs)& $ 4.919^{+0.017}_{-0.02} $& $ 4.92^{+0.017}_{-0.019} $& $ 4.919^{+0.017}_{-0.02} $	\\
	$T_{\rm eff,B}$ (K)& $ 3148\pm31$& $ 3140\pm31$& $ 3148.0\pm31$	\\
	{\bf Eclipse Fitting Parameters} &&&\\
	$D_{\rm pri}$& $ 0.02719^{+0.00027}_{-0.00029} $& $ 0.02721^{+0.00027}_{-0.00028} $& $ 0.02719^{+0.00027}_{-0.00029} $	\\
	$D_{\rm sec}$& $ 0.000716^{+0.000035}_{-0.000031} $& $ 0.000714^{+0.000033}_{-0.000030} $& $ 0.000715^{+0.000033}_{-0.000030} $	\\
	$T_{\rm 14, pri}$ (days) & $0.19655^{+0.00059}_{-0.00056} $& $ 0.19658^{+0.00059}_{-0.00055} $& $ 0.19656^{+0.00059}_{-0.00056} 	$\\
	$k = R_{\rm B}/R_{\rm A}$& $ 0.1827^{+0.0032}_{-0.0027} $& $ 0.1825^{+0.003}_{-0.0026} $& $ 0.1827^{+0.0031}_{-0.0026} $	\\
	$b_{\rm pri} $ & $ 0.9010^{+0.0091}_{-0.0082} $& $ 0.9005^{+0.0086}_{-0.008} $& $ 0.901^{+0.0088}_{-0.0082} $	\\
	$i(^{\circ})$& $ 86.426^{+0.051}_{-0.058} $& $ 86.429^{+0.051}_{-0.055} $& $ 86.426^{+0.051}_{-0.056} $	\\
	$h_1$& $ 0.363^{+0.043}_{-0.044} $& $ 0.367^{+0.043}_{-0.043} $& $ 0.363^{+0.042}_{-0.043} $	\\
	$h_2$& $ 0.206^{+0.044}_{-0.045} $& $ 0.204^{+0.044}_{-0.044} $& $ 0.206^{+0.045}_{-0.045} $	\\
	{\bf Radial Velocity Fitting Parameters} &&&\\
	$K$ (m/s)& $ 21163.2^{+3.5}_{-3.3} $& $ 21163.2^{+3.5}_{-3.3} $& $ 21163.1^{+3.5}_{-3.3} $	\\
	jitter (m/s)& $ 10.6^{+2.8}_{-2.0} $& $10.6^{+2.7}_{-2.0} $& $ 14.4^{+5.3}_{-4.3} $	\\
	{\bf Orbital Parameters} &&&\\
	$T_0$ (BJD)& $ 2457441.916072^{+0.000046}_{-0.000045} $& $ 2457430.524283^{+0.000043}_{-0.000042} $& $ 2457441.916071^{+0.000042}_{-0.000042} $	\\
	$P$ (days)& $ 11.3917809^{+0.0000050}_{-0.0000049} $& $ 11.391781^{+0.0000050}_{-0.0000049} $& $ 11.3917808^{+0.0000049}_{-0.0000049} $	\\
	$a$ (AU)& $ 0.1155^{+0.0019}_{-0.0037} $& $ 0.1164^{+0.00087}_{-0.0019} $& $ 0.115^{+0.0014}_{-0.0013} $	\\
	$a/R_{\rm A}$& $ 14.321^{+0.079}_{-0.087} $& $ 14.327^{+0.077}_{-0.082} $& $ 14.321^{+0.077}_{-0.084} $	\\
	$ e\cos\omega$& $ 0.05604^{+0.00012}_{-0.00013} $& $ 0.05604^{+0.00012}_{-0.00013} $& $ 0.05604^{+0.00013}_{-0.00012} $	\\
	$ e\sin\omega$& $ -0.01234^{+0.00014}_{-0.00015} $& $ -0.01234^{+0.00014}_{-0.00015} $& $ -0.01234^{+0.00014}_{-0.00015} $	\\
	$e$& $ 0.05738^{+0.00012}_{-0.00012} $& $ 0.05738^{+0.00012}_{-0.00012} $& $ 0.05738^{+0.00012}_{-0.00012} $	\\
	$ \omega$ $(^{\circ})$& $ -12.42^{+0.15}_{-0.16} $& $ -12.42^{+0.14}_{-0.16} $& $ -12.42^{+0.14}_{-0.15} $	\\
	\hline
	\end{tabular}
\end{table*}

\newpage

\begin{table*}
	\centering
	\caption{Fitted parameters for the eclipsing single-lined spectroscopic binary EBLM J2217-04 using three different models for the primary star.}
	\label{tab:2217params}
	\begin{tabular}{lccc}
	\hline
	& MIST & YY & Torres	 \\
	\hline
	\hline
	{\bf Primary Star Parameters} &&&\\
	$M_{\rm A} (M_{\odot})$& $ 1.065^{+0.06}_{-0.064} $& $ 1.124^{+0.12}_{-0.07} $& $ 1.17^{+0.082}_{-0.087} $	\\
	$R_{\rm A} (R_{\odot})$& $ 1.569^{+0.029}_{-0.03} $& $ 1.596^{+0.047}_{-0.033} $& $ 1.614^{+0.034}_{-0.037} $	\\
	$\rho_{\rm A}$ (cgs)& $ 0.3883^{+0.0067}_{-0.0065} $& $ 0.39^{+0.0068}_{-0.0069} $& $ 0.3914^{+0.0069}_{-0.0073} $	\\
	$\log g_{\rm A}$ (cgs)& $ 4.074^{+0.011}_{-0.012} $& $ 4.083^{+0.016}_{-0.012} $& $ 4.09^{+0.012}_{-0.015} $ \\
	$T_{\rm eff,A} (K)$& $ 5572.0^{+54.0}_{-52.0} $& $ 5625.0^{+61.0}_{-60.0} $& $ 5565.0^{+59.0}_{-54.0} $	\\
	$\rm [Fe/H]$& $ 0.32^{+0.14}_{-0.23} $& $ 0.37^{+0.2}_{-0.28} $& $ 0.43^{+0.36}_{-0.4} $	\\
	{\bf Secondary Star (M-Dwarf) Parameters} &&&\\
	$M_{\rm B} (M_{\odot})$& $ 0.2233^{+0.0078}_{-0.0085} $& $ 0.231^{+0.014}_{-0.0091} $& $ 0.237^{+0.01}_{-0.011} $	\\
	$R_{\rm B} (R_{\odot})$& $ 0.2387^{+0.0047}_{-0.0047} $& $ 0.2428^{+0.0070}_{-0.0054} $& $ 0.2456^{+0.0054}_{-0.0059} $	\\
	$\rho_{\rm B}$ (cgs)& $ 21.64^{+0.65}_{-0.64} $& $ 21.28^{+0.69}_{-0.69} $& $ 21.08^{+0.7}_{-0.63} $	\\
	$\log g_{\rm B}$ (cgs)& $ 5.0116^{+0.0071}_{-0.0068} $& $ 5.0122^{+0.0069}_{-0.0073} $& $ 5.0122^{+0.0071}_{-0.007} $	\\
	$T_{\rm eff,B}$ (K)& $ 3009\pm22$& $3028\pm29$& $ 3020.\pm44$	\\
	{\bf Eclipse Fitting Parameters} &&&\\
	$D_{\rm pri}$& $ 0.0242^{+0.00016}_{-0.00016} $& $ 0.02418^{+0.00017}_{-0.00017} $& $ 0.0242^{+0.00016}_{-0.00017} $	\\
	$D_{\rm sec}$& $ 0.000502^{+0.000021}_{-0.000021} $& $ 0.000501^{+0.000021}_{-0.000022} $& $ 0.000502^{+0.000021}_{-0.000021} $	\\
	$T_{\rm 1,4, pri}$ (days)& $ 0.19477^{+0.00042}_{-0.00041} $& $ 0.19479^{+0.00042}_{-0.00041} $& $ 0.19475^{+0.00042}_{-0.00041} $	\\
	$k = R_{\rm B}/R_{\rm A}$& $ 0.15557^{+0.00050}_{-0.00053} $& $ 0.1555^{+0.00054}_{-0.00053} $& $ 0.15557^{+0.00051}_{-0.00054} $	\\
	$b $ & $0.7046^{+0.0067}_{-0.0070} $& $ 0.7041^{+0.0069}_{-0.0067} $& $ 0.704^{+0.0069}_{-0.0071} $	\\
	 $i$ $(^{\circ})$& $ 86.457^{+0.053}_{-0.051} $& $ 86.46^{+0.051}_{-0.055} $& $ 86.462^{+0.054}_{-0.054} $	\\
	$h_1$& $ 0.458^{+0.036}_{-0.038} $& $ 0.451^{+0.037}_{-0.039} $& $ 0.466^{+0.039}_{-0.045} $	\\
	$h_2$& $ 0.223^{+0.047}_{-0.046} $& $ 0.233^{+0.05}_{-0.046} $& $ 0.216^{+0.054}_{-0.05} $	\\
	{\bf Radial Velocity Fitting Parameters} &&&\\
	$K$ (m/s)& $ 19936.0^{+13.0}_{-13.0} $& $ 19936.0^{+13.0}_{-12.0} $& $ 19936.0^{+13.0}_{-13.0} $	\\
	jitter (m/s)& $ 41^{+18}_{-15} $& $ 43^{+20}_{-16} $& $ 41^{+19}_{-16} $	\\
	{\bf Orbital Parameters} &&&\\
	$ T_0$ (BJD)& $ 2457009.828256^{+0.000056}_{-0.000056} $& $ 2457009.828249^{+0.000057}_{-0.000057} $& $ 2457009.828242^{+0.000057}_{-0.000056} $	\\
	$P$ (days)& $ 8.1552483^{+0.0000071}_{-0.0000077} $& $ 8.1552484^{+0.0000072}_{-0.0000078} $& $ 8.1552484^{+0.0000072}_{-0.0000076} $	\\
	$a$ (AU)& $ 0.0863^{+0.0015}_{-0.0017} $& $ 0.0877^{+0.0027}_{-0.0017} $& $ 0.0888^{+0.0019}_{-0.0021} $	\\
	$a/R_{\rm A}$& $ 11.821^{+0.063}_{-0.061} $& $ 11.823^{+0.061}_{-0.067} $& $ 11.828^{+0.065}_{-0.065} $	\\
	$e\cos\omega$& $ 0.03117^{+0.00031}_{-0.00031} $& $ 0.03119^{+0.00032}_{-0.0003} $& $ 0.03117^{+0.00032}_{-0.00031} $	\\
	$e\sin\omega$& $ 0.03461^{+0.00072}_{-0.00074} $& $ 0.03463^{+0.00075}_{-0.00076} $& $ 0.0346^{+0.00071}_{-0.00074} $	\\
	$e$& $ 0.04658^{+0.00059}_{-0.00059} $& $ 0.0466^{+0.00064}_{-0.00058} $& $ 0.04657^{+0.00059}_{-0.00059} $	\\
	$\omega$ $(^{\circ})$& $ 47.99^{+0.64}_{-0.66} $& $ 48.0^{+0.65}_{-0.7} $& $ 47.98^{+0.64}_{-0.67} $	\\
	
	\hline
	\end{tabular}
\end{table*}

\newpage

\newpage

\bsp	
\label{lastpage}
\end{document}